# Nanoscale Interlayer Defects in Iron Arsenides

Q. Zheng,[1] M. Chi,[2] M. Ziatdinov,[2] L. Li,[1] P. Maksymovych,[2]
M. F. Chisholm,[1] S. V. Kalinin,[2] A. S. Sefat[1,*]

[1] *Materials Science and Technology Division, Oak Ridge National Laboratory, Oak Ridge, TN 37831*
[2] *Center for Nanophase Materials Sciences, Oak Ridge National Laboratory, Oak Ridge, TN 37831*

[*] *sefata@ornl.gov*

Using a local real-space microscopy probe, we discover evidence of nanoscale interlayer defects along the *c*-crystallographic direction in BaFe$_2$As$_2$ ('122') based iron-arsenide superconductors. We find ordered 122 atomic arrangements within the *ab*-plane, and within regions of ~10 to 20 nm size perpendicular to this plane. While the FeAs substructure is very rigid, Ba ions are relatively weakly bound and can be displaced from the 122, forming stacking faults resulting in the physical separation of the 122 between adjacent ordered domains. The evidence for interlayer defects between the FeAs superconducting planes gives perspective on the minimal connection between interlayer chemical disorder and high-temperature superconductivity. In particular, the Cooper pairs may be finding a way around such localized interlayer defects through a percolative path of the ordered layered 122 lattice that may not affect $T_c$.

## I. INTRODUCTION

Since the discovery of fluorine-doped LaFeAsO with high-temperature superconductivity (HTS) below $T_c$ = 26 K [1], much fundamental research on iron arsenides has continued to expand the understanding of HTS. The well-known $A$Fe$_2$As$_2$ ($A$=Ca, Sr, Ba) parents ('122'), which adopt the tetragonal ThCr$_2$Si$_2$-type structure can become superconducting with only a few percent chemical substitution on $A$ [2], or Fe [3]. Just like other quantum materials, disordered or distorted structures [4-7], vacancies [4, 6, 8-11], inhomogeneities [12-14], and superstructures [6, 9, 10, 15] are shown to exist within iron arsenides and may cause correlated behavior. For example, using aberration-corrected scanning transmission electron microscopy (STEM) imaging within the *ab*-plane there is evidence of lattice distortions due to the shift of iron sub-lattice, in CaFe$_2$As$_2$ that may cause changes in the bulk antiferromagnetic ordering temperature [16], and in cobalt-doped BaFe$_2$As$_2$ resulting in orthorhombicity signatures [7]; additionally, combined STEM and electron energy loss spectroscopy (EELS) data show nanometer Pr atomic clustering in Pr-doped CaFe$_2$As$_2$ resulting in filamentary superconductivity [14].

The microscopic lattice defects have been investigated within the intralayer of *ab*-plane, since HTS is associated with the FeAs layers and because 122 crystals easily cleave perpendicular to *c*. Here and for the first time, we prepare and collect atomic-resolution STEM data on many crystals, along both intralayer [001] and interlayer [100] crystallographic directions. We find that although [001] looks atomically ordered, there are interlayer barium defects along [100]. In order to understand if this is related to the chemical-doping effects of 122s, we compare the atomic structure images of the undoped 122 'parent' with cobalt-doped and gold-doped 122 crystals, and surprisingly find all of them to have similar interlayer defects. These interlayer defects affect the cleavable nature of 122 crystals based on crystallographic covalently-bonded layers of FeAs separated by Ba and may not have great effects on HTS if they are small enough;

they may improve pinning properties leading to higher critical current density. Here we present evidence of microscopic imperfections in the 122 crystals combining optical microscopy, STEM, and EELS data to reveal atomic structure arrangements and chemical compositional variations.

## II. EXPERIMENTAL DETAILS

Crystals of parent BaFe$_2$As$_2$ (122), Ba(Fe$_{0.928}$Co$_{0.072}$)$_2$As$_2$ (Co-122) and Ba(Fe$_{0.928}$Au$_{0.009}$)$_2$As$_2$ (Au-122) were grown using the conventional high-temperature self-flux growth technique, as mentioned in Refs [17, 18]. The crystals are thin flat pieces with thickness of ~0.1 mm along $c$ [001] direction out of each flat crystal face. Optical microscopy (OM) images along [100] direction were obtained on a Leica S6D Greenough stereo microscope. Back-scattered electron images of cleaved surfaces and cross sections of the three crystals were acquired on a Hitachi S3400 Scanning Electron Microscopy (SEM) operated at 15 kV. Scanning tunneling microscopy measurements for Au-122 were carried out using a Joule-Thomson scanning tunneling microscope (JT-STM, Specs, Berlin). A tungsten STM tip was prepared by gentle field emission at a clean Ag(111) sample. The Au-122 crystals were cleaved *in situ* in the STM machine chamber at approximately 110 K, and measured at 77 K, with $U_{sample}$= -100mV, $I_{setpoint}$=1 nA. Thin specimens oriented along the [001] or [100] direction for scanning STEM analyses were prepared by focus ion beam (FIB) milling from BaFe$_2$As$_2$-based crystals, and subsequently thinned using Ar ion milling with liquid nitrogen cooling, in vacuum, and at a weak beam of 1.5 kV and 3mA for 20 min. High-angle annular dark-field (HAADF) imaging and EELS analyses were carried out in an aberration-corrected FEI Titan S 80-300 equipped with a Gatan Image Filter (GIF, Quantum-865) at 300 kV with probe size < 1 Å, or using a VG Microscopes HB603 operated at 300 kV. The probe convergence semi-angle of 30 mrad and an inner collection semi-angle of 65 mrad were used for HAADF imaging. EELS data were collected using a dispersion of 0.25 eV per channel and a collection angle of 40 mrad. Average EEL spectra were acquired for ~ 0.5 nm*2 nm defect and defect-free regions, respectively.

## III. RESULTS & DISCUSSION

Many regions in undoped parent BaFe$_2$As$_2$ (122), and Au-doped (Au-122) and Co-doped (Co-122) crystals were randomly selected to carry out atomic HAADF imaging in [001] projection. All show overall perfect atomic structures without visible defects, as one image exemplifies in Fig. 1a, exhibiting the usual *Z* contrast. The brightest spots are columns of alternating Ba and As atoms, while less intense spots are columns of Fe atoms. Also, we cleaved many crystals to acquire morphology of these planar surfaces, and the surfaces of parent and doped crystals are all shiny and flat when examining with an optical microscope. However, a few of the surface images at micron scales obtained by SEM show steps and isolated island-like surface morphologies, as is seen in Fig. 1b. In fact, further magnification into the nanometer scales by STM in the [001] projection shows small domains, each containing defects with dimensions of several nanometers, as shown in Fig. 1c. Interestingly, an obvious but unexpected lattice shift could also be found between adjacent domains in STM (Fig. 1d). Such small imperfect atomic arrangements within the plane signal the potential for complex stacking structures perpendicular to [001] direction as is studied here. Local-scale analyses using STEM images along [001] have already found structural distortions in Co-122 [7] and parent CaFe$_2$As$_2$ crystals [16] that control magnetic correlations and the potential for superconductivity. Such aberration-corrected STEM images have yielded similar real-space images with resolution on the order of atom sizes in the plane of the sample. However, since HAADF imaging is due to the collective signal of electrons scattered by atomic columns along a certain direction (probing the projected ion-column signals of ~20 nm thickness), the images along [001] would hide details of stacking relationships between the imperfect layers, and the potential causes of such lattice distortions. Therefore, nanoscale investigations perpendicular to *ab* plane i.e., [001] direction, are indispensable to realize the full nature of crystalline arrangement and its potential influence on bulk behaviour. We investigate this next.

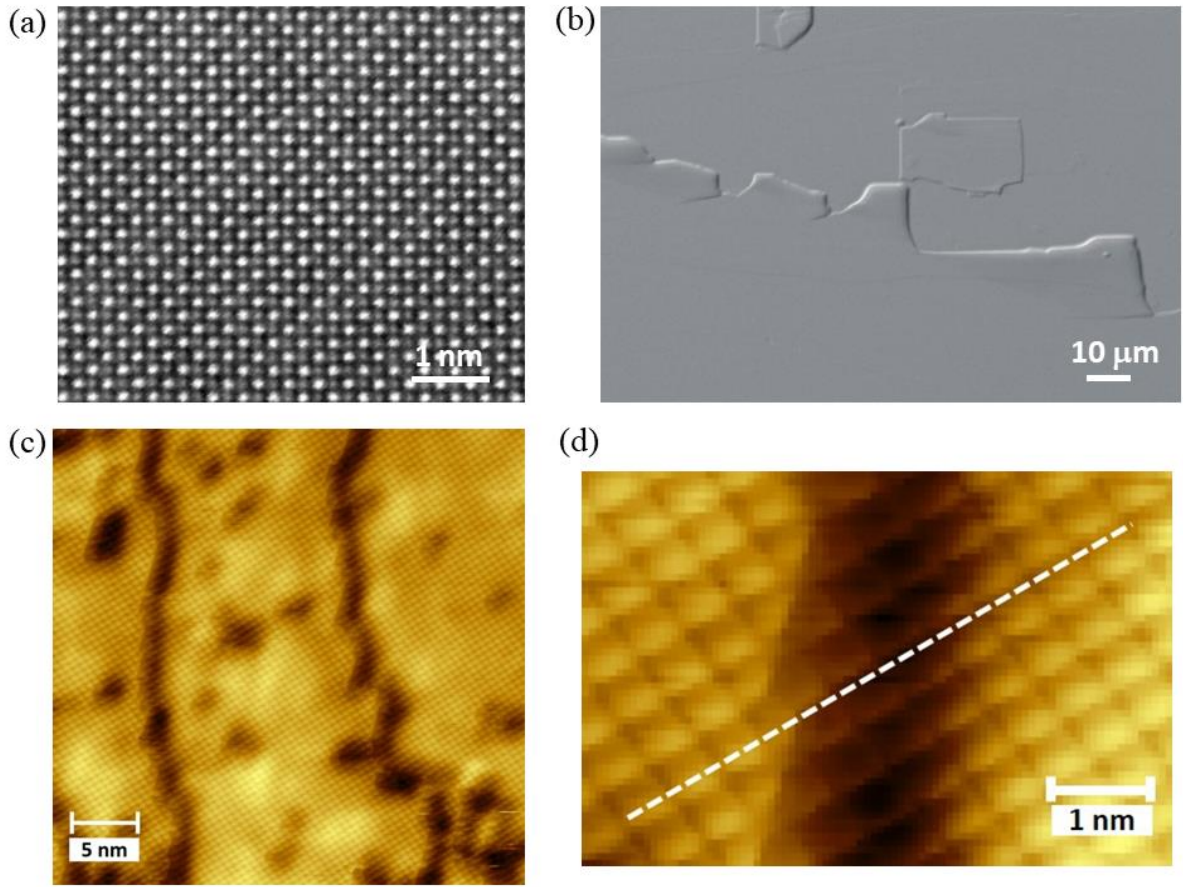

**Figure 1.** Plane-view images of typical 122 iron arsenide in [001] projection. (a) Atomic resolution annular dark-field STEM image; white spots are columns of alternating Ba and As atoms, while gray spots are Fe. (b) SEM and (c) STM images of surface morphology, revealing step-like domains, in one cleaved surface of Au-122 crystal. (d) Atomic resolution STM image of a domain boundary showing a lattice shift between two adjacent domains.

Optical microscopy and SEM were employed to study the 'interlayer' structure perpendicular to the *c*-direction in 122, Au-122 and Co-122 crystals, at the micron scales. As OM images show for the top row of Fig. 2, respectively, comparable amounts of defects are observed in all three types of crystals. The corresponding STEM images, in Fig. 2, reveal more magnified features of these interlayer defects. Either infinite or finite two dimensional (2D) interlayer defects seem to separate the 122 crystalline domains that are of at least several microns thickness. All the images taken for randomly selected 122, Au-122 and Co-122 crystals show these features, which indicate the possible prevalent universality of interlayer defects in flux-grown crystals. At more microscopic levels, as revealed by the low-magnification HAADF image bottom row of Fig. 2, even higher densities of these interlayer defects are visible, separating the crystalline domains into much thinner 2D blocks with thickness of ~10 to 20 nm. Similar to OM and SEM observations, in HAADF images some interlayer defects are continuous across the crystal in view, while some are discontinuous, with lengths sometimes even shorter than 10 nm.

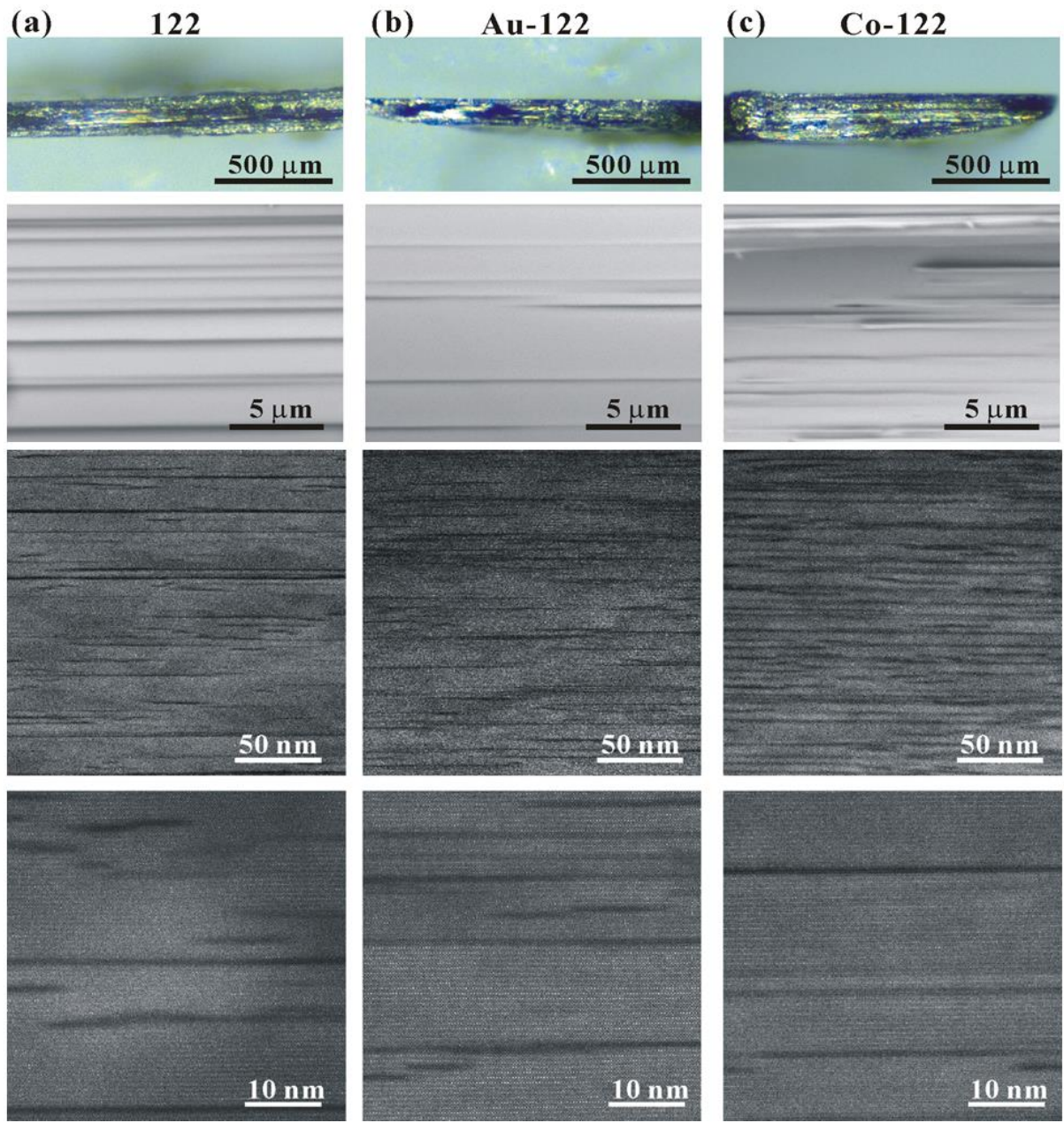

**Figure 2.** Interlayer defects in BaFe$_2$As$_2$-based crystals, as revealed by the micron-scale low-magnification optical microscopy (top panel) and higher magnification HAADF images (rest of panels) perpendicular to the [001] direction for (a) 122, (b) Au-122, and (c) Co-122. Domains of ~10-20 nm can be separated by interlayer defects in the crystals.

Subsequently, the domain regions and the interlayer defects in the parent and Au-122 are further studied by atomic-resolution HAADF imaging. As shown in Fig. 3a, domain regions in 122 crystals can be structurally perfect (as reported before) consistent with the projection of the 122 structure along the [100] direction (inset in this figure). The perfectly arranged pristine atomic areas for 122 and 1111 arsenides and 11 selenides have been reported [19, 20] and used to estimate the local Fe magnetic moment and Fe orbital occupations. Atomic-resolution HAADF-STEM imaging has demonstrated thickness-dependent modulation of intensities within atomic columns (Fe, O and La, As) for [001]-oriented LaFeAsO single crystals [20, 21] with additional speculative reasons for presence of an amorphous layer arising from sample preparation, as well as specimen misorientation with respect to the [001] zone axis. We report here that the observed atomic-column intensity modulations [21], and perhaps the large local fluctuation iron moments [19], and the room-temperature distortions in the crystal structure [7] that are reported in [001] projections may be caused by these interlayer defects that we show in this report, which signal may be averaged through transmission of electrons of several nanometer layers.

The interlayer defects between adjacent perfect domains in 122 crystals are believed to be the result of compositional inhomogeneity and appear as faults in the stacking along the *c*-axis, as is described here. One defect type is stacking faults (marked between two white solid lines in Fig. 3b) that originate from barium deficiency between the FeAs-layers. As one can see, the stacking structure projection in the inset of Fig. 3b, is similar to the FeSe '11' structure for which the interlayer alkali-metal element does not exist. Here, there are three $Fe_2As_2$ layers stacked together without two Ba layers. The Ba layers, marked by two white solid lines, show a transverse shift of half a unit cell along *b*, which is the same as if there was no such stacking fault. If there were no additional $Fe_2As_2$ layers inserted, then one would suspect the distances between the bariums layers to be $1/2c$, however here, the distance between these two marked Ba layers increases to ~2.4 times of $1/2c$. Fig. 3c is another region in this 122 crystal, and the upper interlayer stacking fault is the same with the one in Fig. 3b, revealing that this type of a stacking fault could be commonly found in the 122 type crystals. However, there is another type of stacking fault, shown in the lower part of the image in Fig. 3c: the distance between the Ba layers is nearly the same at 2.3 times of $1/2c$, though there is no shift between Ba atoms between layer planes. Although there seems to be a difference between the two types of stacking faults, and although it is difficult to analyze by spectroscopy, it is clear that a Ba layer gets substituted by an FeAs network. A similar stacking fault is given in Fig. 3d, which also clearly shows the end of one layer: along the white arrow direction, the 2D Ba layer suddenly changes into an $Fe_2As_2$ layer, and this then causes the expansion of Ba-Ba distances to 1.25 *c*, and bending of the domain above it. Moreover, yet another type of stacking fault could be seen in Fig. 3c, where an additional 2D $Fe_2As_2$ layer (marked as an arrow) has been inserted into the structure, resulting in expansion of Ba distances to 1.15 *c*. More interestingly, a 0.2 *a* shift in the Ba layer plane is found between the two sets of domains, above and below the stacking fault. The one end of a stacking fault in Fig. 3e also shows a sudden change of 2D Ba layer into $Fe_2As_2$ layer. Because of such variations of stacking faults, the atomic arrangements cannot be easily obtained by EELS mapping. However, we acquired EELS spectra, as one exemplified in the Fig. 3e inset; it shows more Fe content in the defect regions than that in the perfect domain regions, consistent with atomically resolved stacking faults, due to Ba deficiencies as discussed above. Fig. 3f shows another type of 2D defect, with EELS results on these defect areas to be more Ba-rich. Such defects are not just seen in our single crystals. They are in fact seen in films too: cross-sectional low-resolution TEM images of Ba122:Co thin films with thickness of ~150 nm, on $CaF_2$ substrates [22] or on Fe-buffered MgO template [23].

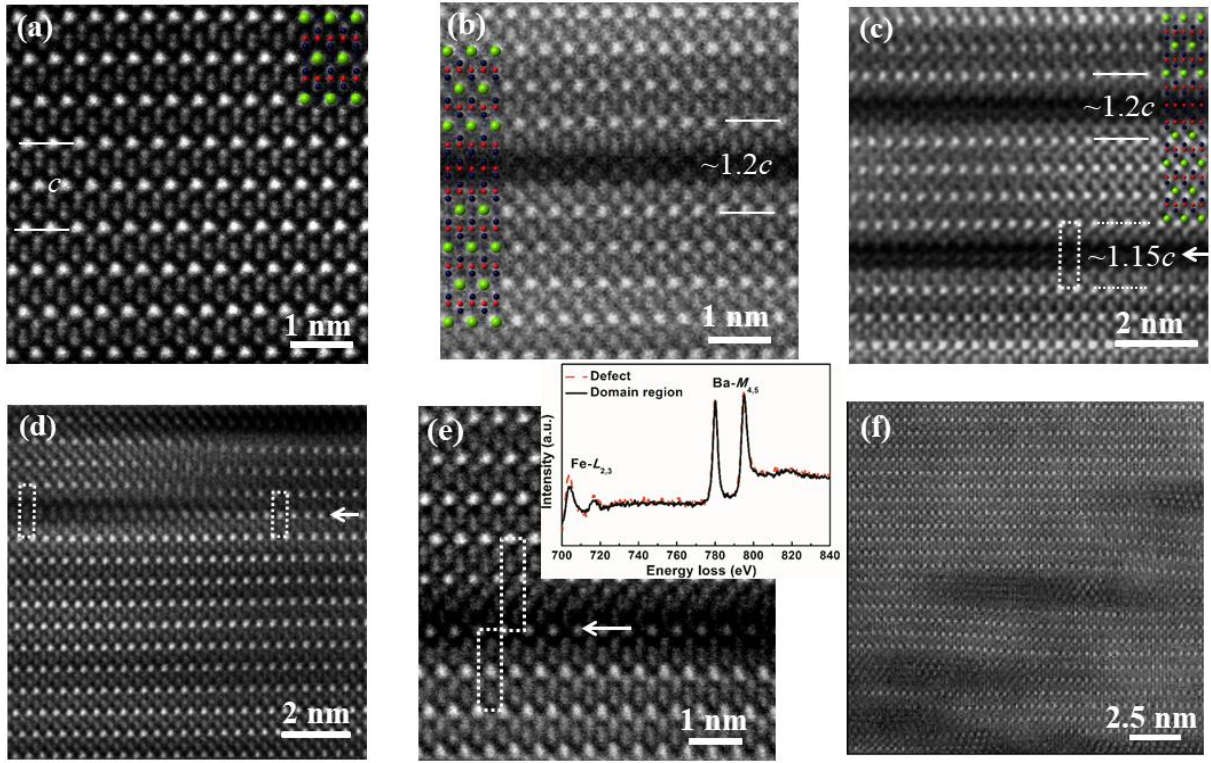

**Figure 3:** Atomic-resolution annular dark-field STEM images showing typical interlayer stacking faults that can be observed in any 122 iron arsenides, along the [100] direction. (a) Perfect atomic-resolution structure in each domain regions; the overlay is the BaFe$_2$As$_2$ structure in the [100] projection (Ba-green, As-black, Fe-red). Here are a few defect structures: (b) shows that two additional FeAs layers can be inserted in a half unit cell; the overlay shows the arrangement of them; (c) displays two type of stacking faults: the top defect is the same with that in (b), while the bottom one is due to substitution of the Ba layer on $c=1/2$ plane by FeAs network; (d) illustrates one end of an interlayer stacking fault. (e) presents an interlayer shift between the two sets of domains, above and below a stacking fault, with EELS spectra of the domain and this defect region. (f) gives Ba rich areas expanding the overall lattice. (a, e) are the results on the Au-122 crystal; (f) is on Co-122 crystal; (b,c,d) are the images of the parent 122.

## IV. CONCLUSIONS

In this study, we used a range of resolution of microscopy techniques from optical microscopy, SEM, STM, to STEM-HAADF, to study the nanoscale interlayer defects in parent 122, Co- and Au-122 crystals. At the micron-scale, obvious interlayer-defects separate the crystalline domains, while nano-scale images reveal smaller domains with thicknesses of 10 to 20 nm along $c$, and variable stacking faults void or rich with barium separating FeAs superconducting layers. These stacking faults could result in separation and expansion of lattices and may even be the cause of strain within 122 crystals. This new evidence of crystalline imperfections in BaFe$_2$As$_2$-based crystals at local scales presents an open question of the effect on chemical disorder between 2D superconducting layers on HTS causing bulk properties in quantum materials.

**REFERENCES**

[1] Y. Kamihara, T. Watanabe, M. Hirano and H. Hosono, *J. Am. Chem. Soc.* **130** (2008), 3296.
[2] M. Rotter, M. Tegel and D. Johrendt, *Phys. Rev. Lett.* **101** (2008), 107006.
[3] A. S. Sefat, R. Jin, M. A. McGuire, B. C. Sales, D. J. Singh and D. Mandrus, *Phys. Rev. Lett.* **101**, (2008), 117004.
[4] Z. Wang, Y. J. Song, H. L. Shi, Z. W. Wang, Z. Chen, H. F. Tian, G. F. Chen, J. G. Guo, H. X. Yang and J. Q. Li, *Phys. Rev. B* **83** (2011), 140505.
[5] R. Cortes-Gil, D. R. Parker, M. J. Pitcher, J. Hadermann and S. J. Clarke, *Chem. Mater.* **22** (2010), 4304.
[6] S. M. Kazakov, A. M. Abakumov, S. González, J. M. Perez-Mato, A. V. Ovchinnikov, M. V. Roslova, A. I. Boltalin, I. V. Morozov, E. V. Antipov and G. Van Tendeloo, *Chem. Mater.* **23** (2011), 4311.
[7] C. Cantoni, M. A. McGuire, B. Saparov, A. F. May, T. Keiber, F. Bridges, A. S. Sefat and B. C. Sales, *Adv. Mater.*, **27** (2015), 2715.
[8] C. Cantoni, A. F. May, J. E. Mitchell, A. S. Sefat, M. A. McGuire and B. C. Sales, *Microsc. Microanal.* **19** (2013), 340.
[9] H. Cao, C. Cantoni, A. F. May, M. A. McGuire, B. C. Chakoumakos, S. J. Pennycook, R. Custelcean, A. S. Sefat and B. C. Sales, *Phys. Rev. B* **85** (2012), 054515.
[10] Z. Wang, Y. Cai, H.-X. Yang, H.-F. Tian, Z.-W. Wang, C. Ma, Z. Chen and J.-Q. Li, *Chin. Phys. B.* **22** (2013), 087409.
[11] R. C. Che, F. Han, C. Y. Liang, X. B. Zhao and H. H. Wen, *Phys. Rev. B* **90** (2014), 104503.
[12] M. P. Allan, T. M. Chuang, F. Massee, Y. Xie, N. Ni, S. L. Bud'ko, G. S. Boebinger, Q. Wang, D. S. Dessau, P. C. Canfield, M. S. Golden and J. C. Davis, *Nat. Phys.* **9** (2013), 220.
[13] W. D. Wise, K. Chatterjee, M. C. Boyer, T. Kondo, T. Takeuchi, H. Ikuta, Z. Xu, J. Wen, G. D. Gu, Y. Wang and E. W. Hudson, *Nat. Phys.* **5** (2009), 213.
[14] K. Gofryk, M. Pan, C. Cantoni, B. Saparov, J. E. Mitchell and A. S. Sefat, *Phys. Rev. Lett.* **112** (2014), 047005.
[15] C. Ma, H. X. Yang, H. F. Tian, H. L. Shi, J. B. Lu, Z. W. Wang, L. J. Zeng, G. F. Chen, N. L. Wang and J. Q. Li, *Phys. Rev. B* **79** (2009), 060506.
[16] B. Saparov, C. Cantoni, M. Pan, T. C. Hogan, W. R. Ii, S. D. Wilson, K. Fritsch, B. D. Gaulin and A. S. Sefat, *Scientific Reports* **4** (2014), 4120.
[17] A. S. Sefat, *Curr. Opin. Solid State Mater. Sci.* **17** (2013), 59.
[18] L. Li, H. Cao, M. A. McGuire, J. S. Kim, G. R. Stewart and A. S. Sefat, *Phys. Rev. B* **92** (2015), 094504.
[19] C. Cantoni, J. E. Mitchell, A. F. May, M. A. McGuire, J.-C. Idrobo, T. Berlijn, E. Dagotto, M. F. Chisholm, W. Zhou, S. J. Pennycook, A. S. Sefat, B. C. Sales, *Adv. Mater.* **26** (2014), 6193.
[20] J. V Zaikina, M. Batuk, A. M. Abakumov, A. Navrotsky, S. M. Kauzlarich, *J. Am. Chem. Soc.* **136** (2014), 16932.
[21] P. K. Suri, J. Yan, D. G. Mandrus, D. J. Flannigan, *J. Phys. Chem. C* **120** (2016), 18931.
[22] P. Yuan, Z. Xu, D. Wang, M. Zhang, J. Li, Y. Ma, *Super. Sci. Tech.* **30** (2017), 025001.
[23] S. Trommler, J. Hanisch, V Matias, R. Huhne, E. Reich, K. Iida, S. Haindl, L. Schultz, B. Holzapfel, *Super. Sci. Tech.* **25** (2012), 084019.

**ACKNOWLEDGEMENT**

This manuscript has been authored by UT-Battelle, LLC under Contract No. DE-AC05-00OR22725 with the U.S. Department of Energy. The United States Government retains and the publisher, by accepting the article for publication, acknowledges that the United States Government retains a non-exclusive, paid-up, irrevocable, world-wide license to publish or reproduce the published form of this manuscript, or allow others to do so, for United States Government purposes. The Department of Energy will provide public access to these results of federally sponsored research in accordance with the DOE Public Access Plan.

The research is primarily supported by the U.S. Department of Energy (DOE), Office of Science, Basic Energy Sciences (BES), Materials Science and Engineering Division. The STM and STEM work was sponsored by the Materials Science & Engineering Division of BES DOE and performed through user projects conducted at Center for Nanophase Materials Sciences (CNMS) that is a DOE Office of Science User Facility.